\documentclass[sigconf, nonacm]{acmart}
\AtBeginDocument{%
  }

\acmISBN{978-1-4503-XXXX-X/2018/06}

\setcopyright{none}
\renewcommand\footnotetextcopyrightpermission[1]{}
\usepackage{amsmath}
\usepackage{algorithmic}
\usepackage{graphicx}
\usepackage{textcomp}
\usepackage{xcolor}
\usepackage{xspace}
\usepackage{booktabs}
\usepackage{multirow}
\usepackage{tcolorbox}
\usepackage{subcaption}
\usepackage{enumitem}
\usepackage{float}
\usepackage{pifont}

\newcommand{\techname}{Sherlock\xspace}
\newtcolorbox[]{findingbox}[1]{
    coltitle=white,
    coltext=black,
    fonttitle=\bfseries,
    title=#1,
    boxrule=2pt,
    arc=3pt,
    left=1pt, right=1pt, top=1pt, bottom=1pt,
    colback=gray!20,
    colframe=black!65,
    before skip=0.3em, 
    after skip=0.3em 
}

\newtcolorbox[]{promptbox}{
    coltext=black,
    boxrule=1pt,
    arc=2pt,
    left=1pt, right=1pt, top=1pt, bottom=1pt,
    colback=gray!20,
    colframe=black!65,
    before skip=0.3em, 
    after skip=0.3em,
    fontupper=\scriptsize
}

\begin{document}

\title{Search-Induced Issues in Web-Augmented LLM Code Generation: Detecting and Repairing Error-Inducing Pages}
\author{Guoqing Wang}
\affiliation{%
  \institution{Peking University}
  \city{Beijing}
  \country{China}}
\email{guoqingwang@stu.pku.edu.cn}

\author{Zeyu Sun}
\affiliation{%
  \institution{Institute of Software, Chinese Academy of Sciences}
  \city{Beijing}
  \country{China}}
\email{zeyu.zys@gmail.com}

\author{Xiaofei Xie}
\affiliation{%
  \institution{Singapore Management University}
  \city{Singapore}
  \country{Singapore}}
\email{xfxie@smu.edu.sg}

\author{Yizhou Chen}
\affiliation{%
  \institution{Peking University}
  \city{Beijing}
  \country{China}}
\email{yizhouchen@stu.pku.edu.cn}

\author{Yanchao Tan}
\affiliation{%
  \institution{Fuzhou University}
  \city{Fuzhou}
  \country{China}}
\email{yctan@fzu.edu.cn}

\author{Yifan Zhao}
\affiliation{%
  \institution{Peking University}
  \city{Beijing}
  \country{China}}
\email{zhaoyifan@stu.pku.edu.cn}

\author{Dan Hao}
\affiliation{%
  \institution{Peking University}
  \city{Beijing}
  \country{China}}
\email{haodan@pku.edu.cn}

\renewcommand{\shortauthors}{Wang et al.}

\begin{abstract}
Web-augmented large language models (LLMs) offer promising capabilities for automatic code generation. However, live web search integration exposes models to unreliable or malicious content, leading to \textbf{Search-Induced Issues (SII)}---a novel failure mode wherein external pages mislead LLMs into producing incorrect code.
This paper presents a comprehensive empirical study demonstrating the prevalence and impact of SII across three commercial search API and six advanced LLMs. Our analysis reveals that all web-augmented LLMs evaluated are vulnerable to SII, with the root causes attributed to either misaligned specifications or flawed code implementations in the searched \textbf{Error-Inducing Pages (EIPs)}.

To address this challenge, we propose \textbf{\techname}, an automated framework designed for LLM service providers to proactively safeguard their web-augmented generation systems at scale. \techname operates as a continuous pipeline that first detects potential SII instances. It then conducts debugging to identify EIPs and pinpoint the root cause within the EIPs, and applies a repair by either annotating misaligned content or replacing erroneous code snippets with evaluated solutions from a trusted source.
Experiments show that \techname can identify  EIPs with an F1-score of up to 95\%, and repair 71\% to 100\% of affected generations across the evaluated models, with modest computational overhead. 
Our findings and framework provide practical guidance for improving the reliability of web-augmented LLM-based code generation systems in real-world software engineering scenarios.

\end{abstract}

%
%



\maketitle

\setlength{\floatsep}{4pt plus 4pt minus 1pt}
\setlength{\textfloatsep}{4pt plus 2pt minus 2pt}
\setlength{\intextsep}{4pt plus 2pt minus 2pt}
\setlength{\dbltextfloatsep}{3pt plus 2pt minus 1pt}
\setlength{\dblfloatsep}{3pt plus 2pt minus 1pt}
\setlength{\abovecaptionskip}{3pt}
\setlength{\belowcaptionskip}{2pt}

\section{Introduction}
\label{sec:intro}
Large language models (LLMs) have become foundational technologies across numerous domains, particularly for automated code generation in software engineering (SE)~\cite{chen2021codex, llama, gpt4o}. This new paradigm involves two stakeholders: end-user developers, who leverage LLMs to accelerate work, and LLM service providers (e.g., OpenAI, Google), who build and maintain the platforms. From a developer's perspective, a core limitation of LLMs is their reliance on static, pre-training data, rendering them oblivious to new APIs, libraries, or evolving best practices~\cite{brown2020language, openai2023gpt4, deepseekai2024deepseekv3technicalreport}. To meet this critical need of developers, LLM service providers have widely integrated live web search into their generation pipelines~\cite{chatgpt-search, claude-3.7, gemini-2.5, deepseek-v3}, now standard across all flagship models~\cite{techsur-llm-trends-2025, ms365-copilot-web-access}. However, for providers, this web-integrated approach, while essential for user satisfaction, poses a serious challenge to service reliability and platform trustworthiness, as they are now responsible for mitigating the risks from the vast, unpredictable, and often erroneous content of the open web.

This challenge for LLM service providers is not theoretical. A report on X~\cite{x_case_2024} provides a stark case of this service-level risk. In the incident, a user requests a code script from \texttt{ChatGPT} for purchasing \textit{Solana} tokens via the \textit{pump.fun} portal. In response, the LLM service's automated web search module, attempting to fulfill the request with the most relevant information, identifies and ingests content from \textit{docs.solanaapis.com}—a website that, unknown to the system, contains malicious code. The service then feeds this untrusted content directly into its generation model, which dutifully synthesizes it into a functional but malicious script that exposes the user's private key (see Fig.\ref{Motivating}). The resulting financial loss represents a critical service failure for the provider: the platform intends to be a helpful assistant, but instead delivers a harmful payload.

This failure scenario exemplifies what we define as a Search-Induced Issue (SII): a code generation failure where an LLM produces misleading, incorrect, or harmful code as a direct consequence of incorporating flawed information from an external web page. 
The root cause of this particular SII is the misleading information in the specific web page that the LLM trusted. We term such a problematic source an Error-Inducing Page (EIP): a web page containing outdated, incorrect, or malicious content that misleads an LLM into generating faulty code. While this case serves as a stark warning, it remains unclear whether it represents a rare anomaly or a widespread phenomenon. To move beyond anecdotal evidence, we conduct a preliminary study to quantify the prevalence of SIIs and analyze the underlying characteristics of the EIPs that cause them. From an LLM service provider's perspective, it is a critical assessment of a platform-wide vulnerability.

\begin{figure}[t]
\centering
\includegraphics[width=0.95\linewidth]{figs/motivation_FSE26.png}
\caption{A motivating example of failure caused by SII}
\label{Motivating}
\end{figure}

In our preliminary study, we evaluate LLM-generated code across a broad suite of programming problems under two configurations (i.e., with and without web search). 
Our experimental results demonstrate that, to varying degrees, all web-augmented LLMs evaluated are suffering from SII. 
On average, for every additional correct answer due to the web search content, 73\% additional incorrect answers are also introduced.
To understand the root cause of SII in depth, two authors manually analyze the instances of faulty code and their corresponding searched web pages. This in-depth analysis reveals not only the underlying failure types—incorrect implementations (accounting for 12\% of EIPs) and misaligned specifications (88\%)—but also a critical strategic insight. We find that the impact of these EIPs is significantly amplified when they concern fundamental, reusable code building blocks, such as list operations, common API calls, or standard algorithms.
These foundational functions serve as the essential building blocks for more complex software systems.
This leads to a crucial realization: while tackling SII for every arbitrary, open-ended query is intractable due to the lack of a universal standard for correctness, a more scalable and high-impact strategy is to proactively purify the information sources for foundational, evaluable code snippets, thereby protecting the building blocks of countless future generations.

For LLM service providers, establishing a high-quality ``Correctness Evaluator'' for these foundational tasks is feasible. This Evaluator is not limited to traditional test oracles; it can be instantiated through various industrial-scale methods, such as consistency checks across multiple generation runs, execution against extensive internal regression test suites, or validation against human-written gold standard solutions. By leveraging such an evaluator, service providers can systematically cleanse the ecosystem of fundamental EIPs, enhancing the reliability of the entire generation pipeline and thereby improving the quality of countless future code generations.

This insight shifts our focus from identifying SII to building a targeted and scalable defense for LLM service providers. We introduce \textbf{\techname}, an automated end-to-end pipeline that proactively mitigates SII through cache-level intervention. 
Unlike reactive per-query mitigations, such as repairing individual outputs or filtering retrieved information at inference time, \techname addresses the root cause by repairing the shared web page cache itself. It follows a ``fix-once, protect-many'' strategy: once an EIP is repaired, the threat is neutralized for all subsequent queries that rely on that page.
\techname consists of three stages: (1) \textbf{SII Detection}, which identifies potential SII cases by comparing web-augmented generations against baseline generations verified as correct by the Correctness Evaluator; (2) \textbf{EIP Debugging}, which locates the responsible EIPs and diagnoses their root causes; and (3) \textbf{Cache Repair}, which repairs and updates the cached page with evaluable content verified by the Correctness Evaluator, thereby protecting future queries from the same source-level threat.

To assess the effectiveness of \techname, we conduct an evaluation on six mainstream LLMs across five benchmarks.
The experimental results demonstrate that: (1) \techname can effectively identify EIPs with an F1-score up to 95\%, and diagnose with an accuracy of 90.8\%. (2) For generation errors caused by EIPs, \techname successfully repairs 71\% to 100\% of the affected cases. Under the same evaluation protocol, \techname achieves up to 80.1\% higher macro-averaged repair success rate than representative reactive RAG baselines, while also being complementary to them. (3) In terms of efficiency, \techname requires only 2.95 seconds on average to fix one SII instance.

We summarize the contributions of this paper as follows:

$\bullet$ We are the first to systematically study the impact of SIIs in code generation scenarios and perform an in-depth analysis of EIPs to identify their underlying causes in triggering generation errors. 

$\bullet$ We further propose a proactive framework \textbf{\techname} to mitigate SII for LLM service providers, which automatically debugs and repairs EIPs to ensure correct generations.

$\bullet$ We conduct a comprehensive evaluation to evaluate the effectiveness of \techname.

\section{Related Work}
\label{sec:related_work}

Automatic code generation~\cite{chen2021codex,wang2021codet5} has rapidly advanced with the rise of LLMs~\cite{gpt4o, claude-3.7, gemini-2.5, deepseek-v3, deepseek-r1} trained on massive code corpora, such as Codex~\cite{chen2021codex} and ChatGPT~\cite{gpt4o}. These models achieve impressive performance across standard programming benchmarks. 
Some works~\cite{jain2023llm, luo2023wizardcoder, phind-codellama-34b-v2} further boost the code generation ability through Supervised Fine-Tuning (SFT).
In addition to scaling model size and data diversity, researchers also leverage specialized prompting strategies~\cite{liu2024large, fan2023large, chen2025deep, kim2024language}, such as chain-of-thought (CoT) reasoning~\cite{wei2022chain}, expert prompting~\cite{xu2023expertprompting}, multi-agent collaboration~\cite{huang2023agentcoder, dong2024self}, self-refinement~\cite{kim2024language}, and execution feedback information~\cite{zhong2024debug, huang2023agentcoder}, to further boost the reliability of generated code. 
Despite these advances, the static nature of LLM pre-training poses inherent limitations that models may hallucinate plausible yet incorrect code~\cite{deepseekai2024deepseekv3technicalreport, brown2020language,openai2023gpt4}. These challenges motivate approaches that generate content based on external sources.

Retrieval-Augmented Generation (RAG)~\cite{karpukhin2020dense, lewis2020retrieval, borgeaud2022improving} enhances LLMs by supplementing inputs with relevant retrieved documents from a structured or curated corpus. In the SE context, this typically involves mining codebases, API documentation, or QA forums~\cite{thoppilan2022lamda, sun2024source}. Many prompt-based techniques~\cite{pan2024lost, yang2024exploring} provide LLMs with few-shot retrieval content to enhance LLMs' capabilities. 
Recently, advanced LLMs extend RAG by leveraging live web search~\cite{chatgpt-search,gemini-2.5, perplexity-ai}, enabling LLMs to access broad online resources.
This strategy bridges the gap between LLM knowledge and the rapidly evolving software landscape, yielding improvements in relevance, recency, and factual accuracy. 

However, augmenting LLMs with external knowledge also introduces new risks. Some research focuses on attacking LLMs, such as prompt injection~\cite{yang2024tapi,zhang2024hijackrag}, data poisoning~\cite{zhang2024human, lin2025exploring, zou2024poisonedrag}, and adversarial retrieval~\cite{xue2024badrag}. 
These studies demonstrate the vulnerability of RAG pipelines. Other research also recognizes the inherent limitations of RAG systems, such as their retrieval mechanisms and the issue of low-quality content~\cite{xiang2024certifiably, hwang2024retrieval}.
Hence, researchers propose some defense mechanisms~\cite{zhang2024attacks, zhang2025traceback} and optimization techniques~\cite{xiang2024certifiably, hwang2024retrieval, hwang2025retrieval} to mitigate these problems. For instance, RobustRAG~\cite{xiang2024certifiably} modifies the generation process to make RAG systems less susceptible. These approaches act as a crucial runtime safeguard, enhancing the reliability of individual generation instances.

In contrast, proactive data curation can address the problem at its source: the external knowledge base itself, which has not received much attention before. This is the category where our work, \techname, makes its primary contribution. 
Instead of mitigating the effects of flawed data during each query, \techname is designed as an offline framework to proactively and continuously detect, debug, and repair the flawed web pages before they are ever served to the RAG system. It functions as a sanitization layer for the web-based knowledge source, aiming to improve the overall quality and trustworthiness of the data that RAG systems rely upon.

Therefore, our work is not a competing alternative to in-process defenses but rather an orthogonal and complementary approach. \techname can be deployed by LLM service providers to systematically enhance their web knowledge corpus, reducing the likelihood of SII from the outset. Runtime defenses can then serve as a valuable second line of defense, handling any residual errors or defects from sources not yet curated by our framework.

\section{Preliminary Study}
\label{sec:study}
\subsection{Empirical Design}
In this work, we focus on the problem we term the \textbf{Search-Induced Issue (SII)}.
To address this from the perspective of LLM service providers, we first analyze how prevalent SII is and investigate its root causes, specifically, what characteristics define an EIP.

To assess its prevalence, we perform an extensive preliminary study on advanced LLMs. We generate solutions for every programming problem under two settings (with web search and without web search). By comparing outputs across these settings, we quantify the impact of web search on code correctness.
After that, to uncover root causes, two authors manually inspect each case where only the web‐augmented model fails. For every such instance, we trace the searched pages and label the underlying issue.

This preliminary study is driven by two research questions:
\ding{172} \textbf{P-RQ1:} How widespread is the SII among advanced LLMs when augmented with web search?
\ding{173}  \textbf{P-RQ2:} What is the root cause of LLMs' code generation error caused by web pages?

\noindent \textbf{Experimental LLMs.}
\label{subsec:evaluated_llm}
Since commercial LLMs' search engines, as a core business asset, are completely confidential and not publicly available,  we evaluate the public official web-enabled models through API calling(\texttt{GPT Search}, \texttt{Claude Search}, and \texttt{Gemini Search}).  
However, as explicitly mentioned in their official documentations~\cite{openai-tools-web-search, claude-tools-web-search, gemini-tools-web-search}, they only expose the model-processed answers and some of the referenced URLs. It is not possible to obtain the complete, original search result text and provide references to indirect citations. This black-box nature of commercial LLMs is not conducive to detailed analysis of EIPs.

Hence, for a broader, systematic, and reproducible analysis, we evaluate six mainstream LLMs based on open-source web-augmented implementation, spanning a range of architectural designs: two non-reasoning models (\texttt{GPT-4o}~\cite{gpt4o}, \texttt{DeepSeek-V3}~\cite{deepseek-v3}), two hybrid reasoning models (\texttt{Claude 3.7 Sonnet}~\cite{claude-3.7}, \texttt{Gemini 2.5 Flash}~\cite{gemini-2.5}), and two reasoning-focused models (\texttt{o4-mini}~\cite{o4-mini}, \texttt{DeepSeek-R1}~\cite{deepseek-r1}). These models are chosen for their performance, accessibility, API cost~\cite{openai_pricing}, and representative coverage of current techniques.

\noindent \textbf{Implementation.}
For evaluation, we use \textit{MBPP+}~\cite{evalplus}, a widely adopted benchmark in prior code-generation studies~\cite{dong2024self, huang2023agentcoder, zhong2024debug}. To reduce randomness, each LLM is run three times per programming task with temperature set to 0, following common practice in code generation~\cite{evalplus, deepseek-v3, zhong2024debug}. We also use the maximum available \textit{max\_token} setting for each model to avoid length-related performance constraints~\cite{han2024token}. To reduce prompt-related variance, we directly adopt the released code and prompts from prior work.

For web search, our study comprises two settings. For models with built-in search APIs, we use their official integrated web search functions with default recommended configurations~\cite{openai-tools-web-search, claude-tools-web-search, gemini-tools-web-search}. 
For the other six LLMs, we build a unified and reproducible search environment using the open-source \texttt{LangChain Search} pipeline~\cite{Elovic_gpt-researcher_2023, Chase_LangChain_2022, xiang2024certifiably}, which has been widely used in prior research and agent frameworks~\cite{xue2024badrag, xiang2024certifiably, Chuanhu_and_MZhao_and_Keldos_Chuanhu_Chat_2023, Elovic_gpt-researcher_2023}. As the search engine, we use \textit{Google Search}~\cite{google-custom-search-api}, given its widespread use and its adoption in existing LLM search tools~\cite{web-search-with-google-api, gemini-tools-web-search, claude-tools-web-search}. Although using a unified search engine may introduce some bias, it provides a controllable setup for systematic academic evaluation of closed-source systems.
Following standard RAG practice, we use Top-3 retrieval in the main setting to balance search quality and cost. Since Top-3, Top-5, and Top-10 are all common retrieval settings in prior work~\cite{pan2024lost, yang2024exploring, xiang2024certifiably, hwang2024retrieval, hwang2025retrieval}, we further extend the evaluation to Top-5 and Top-10 in Section~\ref{sec:evalution}.

\noindent \textbf{Evaluation Metric.}
To evaluate the impact of web search, we requires a Correctness Evaluator to establish ground truth and evaluate the correctness. As introduced in Section~\ref{sec:intro}, this evaluator can be instantiated in various ways. For the experiments in this preliminary study, we instantiate this evaluator using the comprehensive test suites provided with the MBPP+ benchmark. We then generate code solutions under two settings for each problem: (1) with web search, where the model incorporates searched web content, and (2) without web search, where the model relies solely on its pre-trained knowledge.
For both settings, we independently generate solutions three times per task. Correctness is determined as follows: (1) correct: a task is considered correct only if all three generated solutions pass the test suite; (2) incorrect: a task is considered incorrect if any of the three solutions fail on the test suite.
This conservative evaluation protocol ensures that reported differences are attributable to the generation setting, not random model fluctuations.

\noindent \textit{New Errors Caused by EIP:}  
We specifically count the number of tasks $E$ for which solutions are correct without web search but become incorrect when web search is enabled. This metric quantifies the negative impact of EIPs on LLMs' code generation performance.

\noindent \textit{Negative Impact Ratio (NIR):}  
To assess the risk-benefit trade-off of web search integration, we define the Negative Impact Ratio as $\text{NIR} = \frac{\text{E}}{\text{C}}$, where $\text{E}$ is the number of new errors caused by EIPs, and $\text{C}$ is the number of problems for which web search enables correct solutions that the model cannot solve without web access. When $\text{C}= 0$, we record it as NA and count \text{E} separately. A lower NIR indicates a more favorable balance, meaning the benefits of web search outweigh its risks.

\noindent \textbf{Pass@1:} Following prior work~\cite{zhong2024debug, evalplus, huang2023agentcoder}, we further report Pass@1~\cite{chen2021codex}. Note that Pass@1 is reported as an auxiliary \emph{generation-level} metric, whereas ``New Correct via Search'', ``New Errors via Search'', and NIR are \emph{conservative task-level} transition metrics computed from three independent generations per task under the all-pass / any-fail protocol described above.

\begin{table}[t]
       \caption{Prevalence and impact of SII across LLMs on MBPP+.}

    \label{empirical_results}
    \setlength{\tabcolsep}{1.5pt}
     \footnotesize
    \begin{tabular}{lccccc}
    \toprule
     Evaluated LLM  & Pass@1 & $\text{Pass@1}_{search}$ & New Correct & New Errors & NIR\\ \midrule
     GPT-Search  & 73.81\% & 75.40\% & 23 & 17 & 73.91\%\\
     Claude-Search & 74.69\% & 79.89\% & 32 & 12 & 37.50\%\\
     Gemini-Search & 77.64\% & 80.16\% & 35 & 25 & 71.43\%\\
     \midrule
    GPT-4o  & 73.81\% & 74.62\% &  32 &  25  &  78.13\%  \\
    DeepSeek-V3 & 77.51\% & 78.83\% & 17 & 16 & 94.12\% \\
    Claude 3.7 Sonnet  & 74.69\% & 77.57\% & 34 & 11 & 32.35\%  \\
    Gemini 2.5 Flash   & 77.64\% & 79.36\% & 32 & 25 & 78.13\% \\
    o4-mini & 75.47\% & 77.08\% &  48 & 44  &  91.67\%   \\ 
    DeepSeek-R1 & 81.48\% & 82.54\%  &  15 & 15 & 100.00\% \\

    \bottomrule
    \end{tabular}
    
\end{table}

\subsection{P-RQ1: Impact of Search-Induced Issue }
To understand the impact of integrating web search, we first aim to confirm that SII is a tangible issue in live commercial systems. 
Table~\ref{empirical_results} reports two complementary views of this impact. Columns 2--3 present Pass@1 as an auxiliary \emph{generation-level} metric, while Columns 4--6 report ``New Correct via Search'', ``New Errors via Search'', and NIR as \emph{conservative task-level} transition metrics computed from three independent generations per task.

The results show that all evaluated commercial LLMs suffer from SII to varying degrees. Although enabling web search generally improves Pass@1, it also introduces new failures. For example, \texttt{GPT-Search} improves Pass@1 to 75.40\%, but while 23 previously failed tasks become correct, 17 previously correct tasks turn incorrect, yielding an NIR of 73.91\%. \texttt{Claude-Search} shows the lowest NIR at 37.50\%, whereas the other two exhibit substantially higher NIRs. These results confirm that SII is a real-world phenomenon and that web search introduces a clear risk--benefit trade-off.

A similar phenomenon can also be found in our open-source web-augmented implementation. For example, \texttt{Gemini 2.5 Flash} produces 25 new incorrect solutions due to SII. 
The similar trends and numerical results indirectly confirm that this widely used open-source implementation has similar trends to the closed-source implementation. 
The reasoning model \texttt{o4-mini} achieves the most newly solved tasks (48), but also the most new failures (44), suggesting that deeper use of searched context may increase vulnerability when the retrieved content is flawed. \texttt{DeepSeek-R1}, \texttt{DeepSeek-V3}, and \texttt{o4-mini} all exhibit high NIRs, indicating that the gains from search can be largely offset by newly introduced errors.

The often high NIRs demonstrate that simply enabling web search is not a universally net-positive strategy. From the perspective of an LLM service provider, this strongly motivates the need for frameworks or mechanisms that can detect and repair EIPs, which is the core objective of our proposed framework \techname.

\subsection{P-RQ2: Error-Inducing Page Types }
While this result in Table~\ref{empirical_results} confirms the problem's existence, the black-box nature~\cite{openai-tools-web-search} of commercial APIs makes a deep, comparative analysis impossible. Therefore, we conduct a broader study in our controlled search setting. 
To understand the underlying causes of errors introduced by web search integration, we further conduct an in-depth manual analysis of the searched web pages involved in the erroneous outputs. We select \texttt{GPT-4o} and \texttt{DeepSeek-V3}'s generations as a representative because both are widely-used in previous work~\cite{zhong2024debug, huang2023agentcoder, gao2023makes, dong2024self, sun2024source}. Our analysis focuses on the characteristics of the web pages searched and utilized by the LLM.

To quantitatively assess these factors, two authors of this paper manually review and annotate each web page utilized by \texttt{GPT-4o} and \texttt{DeepSeek-V3} in the 25/15 instances where it produces incorrect code with web search enabled. After removing duplicate pages, a total of 93 web pages are associated with these failure cases. 
According to specification alignment and implementation correctness, web pages are classified into three categories, i.e., (1) pages of correct implementation and aligned specification with the input programming problem (not EIPs), (2) pages of aligned specification but incorrect implementation (i.e., failing the benchmark's test suite, our instantiated Correctness Evaluator), and (3) pages of misaligned specification. For these pages of misaligned specification, implementation correctness is not further assessed, as we regard that misaligned content should not be directly incorporated by the LLM, irrespective of its intrinsic correctness for its original context.
Disagreements in annotation are resolved through face-to-face discussion. The inter-rater reliability of the annotation process, measured using Cohen’s Kappa~\cite{cohen1960coefficient}, is $> 0.9$, indicating a high degree of agreement and confirming the reliability of our labeling.

\begin{figure}[t]
\centering

\begin{subfigure}{0.98\columnwidth}
    \centering
    \includegraphics[width=\linewidth]{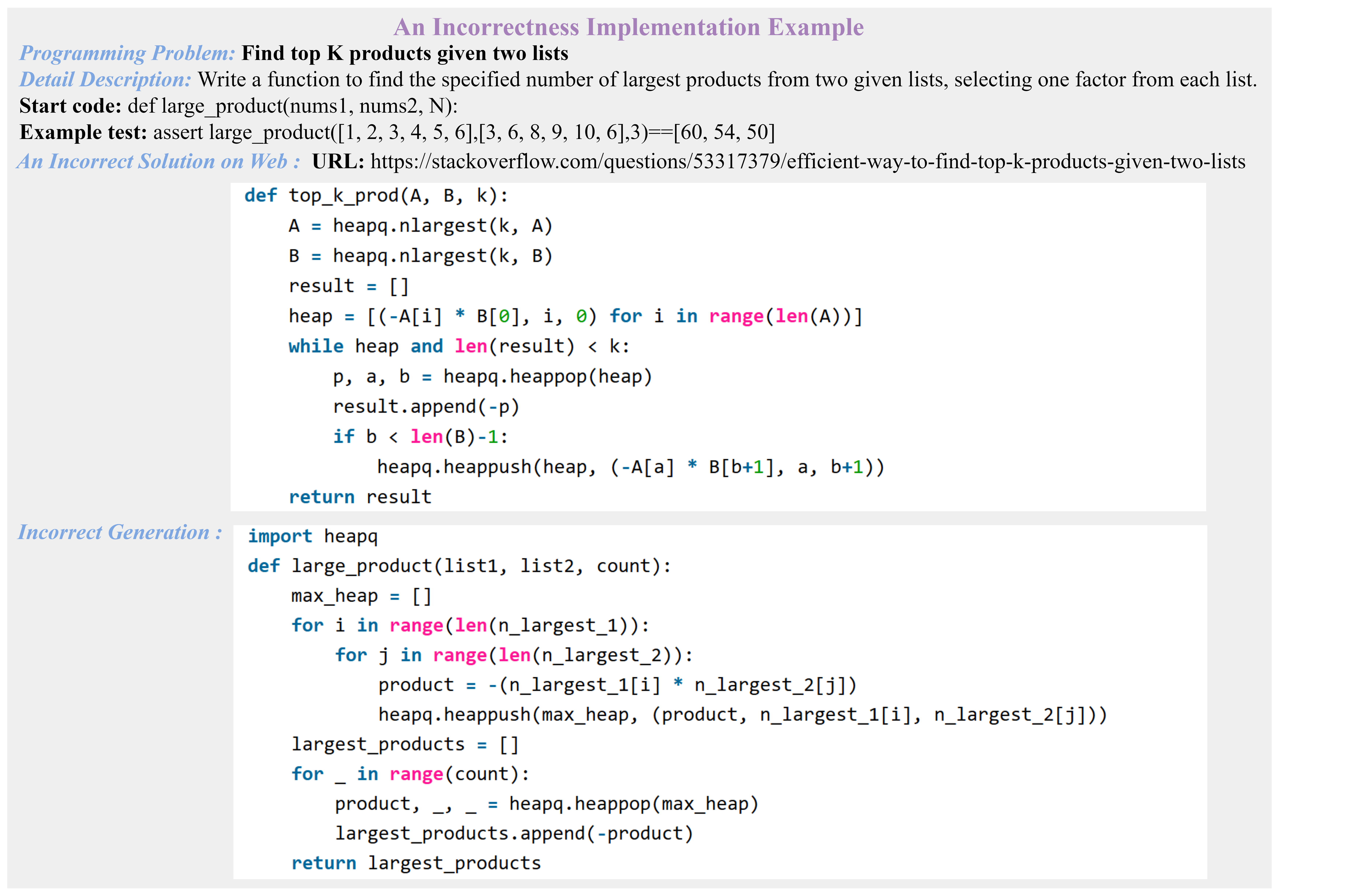}
    \caption{Web Page of Implementation Incorrectness}
    \label{fig:sub1}
\end{subfigure}

\begin{subfigure}{0.98\columnwidth}
    \centering
    \includegraphics[width=\linewidth]{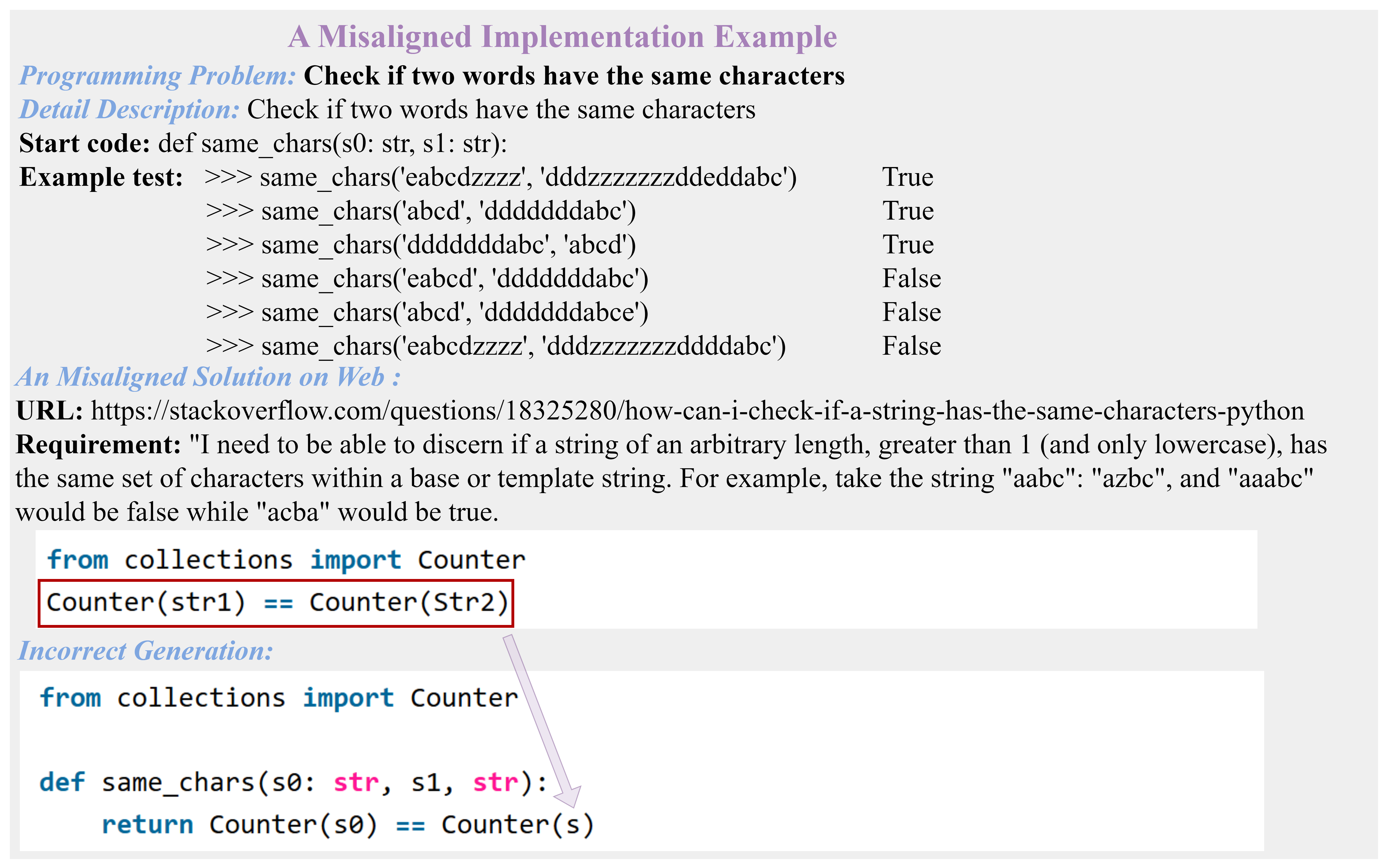}
    \caption{Web Page of Specification Misalignment}
    \label{fig:sub2}
\end{subfigure}
\caption{Two types of EIPs identified in preliminary study.}
\label{PSExample}
\end{figure}

The results of the quantitative analysis for the 93 web pages yield the following distribution: (1) 20 pages (21.5\%) exhibit both correct specification alignment and correct implementation;
(2) 9 pages (9.7\%) have correct specification alignment but contain incorrect implementations;
(3) 64 pages (68.8\%) are misaligned with the problem specification.

A crucial observation from this analysis is that, for the 40 failure instances, there is not a single instance where all three searched web pages fully comply with the specification for the given problem and contain a correct implementation. This suggests that providing relevant and correct search content can significantly mitigate the risk of LLM error. Conversely, the presence of even a single misaligned web page poses a substantial risk of leading the web-augmented LLMs to generate faulty code. To be specific, we observe two scenarios leading to incorrect code generation.

\noindent \textbf{Implementation Incorrectness.} 
The searched web page meets the requirements of the input problem (i.e., specification alignment), but the provided code snippet contains errors or is suboptimal. The LLM, incorporating this flawed implementation, may then reproduce or propagate these errors. As illustrated in Fig.\ref{PSExample}.a, the implementation of \textit{Find top K products given two lists} in the webpage is wrong (negative numbers are not considered). However, the LLM does not recognize the error and refers to the implementation in the web page, resulting in incorrect code generation. 

\noindent \textbf{Specification Misalignment.} 
The searched web page, despite containing correct code for the problem it addresses, is not aligned with the requirements of the input programming problem. Direct use of such content by the LLM can lead to solutions that are incorrect for the given task. As shown in Fig.\ref{PSExample}.b, the implementation of \textit{Check if two words have the same characters} in the webpage is correct, but the detailed requirements of the two programming problems are inconsistent. The problem in the webpage needs to further consider repeated characters, but the input problem does not. The LLM does not distinguish the difference between the two programming problems and directly refers to the implementation, resulting in incorrect code generation.

These findings demonstrate that robust code generation with web-augmented LLMs is contingent on both precise specification matching and high-quality, error-free implementations in retrieved content. 
Misaligned specifications with inputs or incorrect implementation in web pages are the root causes of SII caused by EIP. About 68.8\% of all searched web pages are misaligned with the given specifications, and 9.7\% contain incorrect implementations.

\section{Proposed Approach}
Based on these observations, we propose \techname, which strategically focuses on evaluable software engineering tasks. We design it as a tool for LLM service providers, aiming to build a cache of sanitized and evaluated web content about fundamental programming knowledge. 
We believe this targeted approach is a very pragmatic and impactful approach to addressing current SII challenges.

\subsection{Overview}
\techname is designed to be integrated into an LLM service provider's continuous quality assurance pipeline. Its operation can be conceptualized in two main parts: a high-level SII detection loop and the \techname engine, which is activated by the loop.

\begin{figure*}[]
\centering
\includegraphics[width=\linewidth]{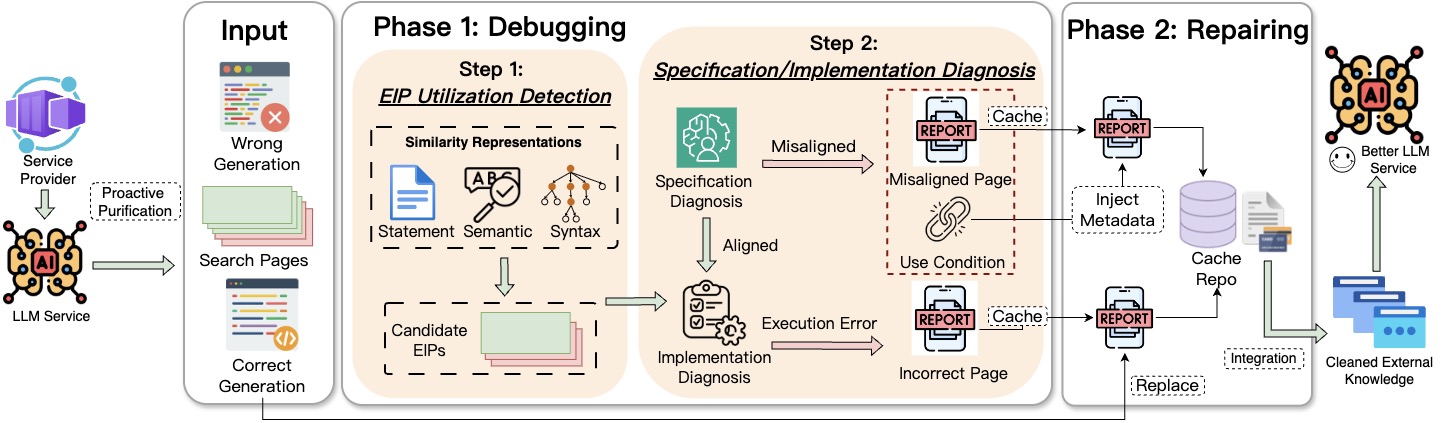}
\caption{An overview of \techname's automated pipeline for LLM service providers.}

\label{overview}
\end{figure*}

The SII detection loop is the system-level process described in Section~\ref {sec:intro}, which continuously compares the outputs of web-augmented generations against their corresponding baseline model generations to detect potential SII instances at scale. When an SII instance is detected, it is passed to the \techname engine, along with the erroneous output and the set of searched web pages.
The \techname engine, as shown in Fig.\ref{overview}, executes a two-phase process on a specific SII instance: a Debugging Phase followed by a Repairing Phase. The Debugging Phase consists of two steps: EIP Utilization Detection to identify candidate EIPs, and Specification/Implementation Diagnosis to pinpoint the root cause within EIPs.
The Repairing Phase leverages two strategies to repair EIPs according to the specification alignment and implementation correctness.
We will detail the inner workings of \techname in the following subsections.

\subsection{Debugging Phase}

In this phase, \techname aims to identify which searched pages are responsible for faulty code generation through a two-step procedure and pinpoint the root causes.
In EIP Utilization Detection, \techname identifies exactly which web pages the LLMs utilize when producing their erroneous output through multi-granular content matching technology. 
In Specification/Implementation Diagnosis, for each utilized page, \techname conducts the specification alignment check and implementation correctness check. Pages that fail either check are output as EIPs and will be further repaired. 

\noindent \textbf{EIP Utilization Detection.}
\techname detects the utilization of web content from multiple perspectives to identify whether code from EIPs is incorporated into the generations. To comprehensively consider the statement, syntax, and semantics, \techname operates at code snippet-level and line-level granularities and leverages several complementary similarity representations, including code text, keywords, abstract syntax trees (AST)~\cite{aho2006compilers}, and data flow~\cite{kildall1973unified}.

The detail detection process is as follows. At the snippet level, \techname extracts all code segments from each web page using an \textit{HTML} parser. For each code snippet, to capture surface-level similarity, \techname evaluates its similarity to the erroneous output using token and textual similarity. Furthermore, to account for possible variable renaming and code restructuring, and capture deeper semantic correspondence, \techname calculates the syntax (AST) similarity and data flow matching.
At the line level, each code snippet is further segmented into individual lines. \techname then compares each line against the error output, focusing primarily on token-level similarity. This fine-grained analysis aims to detect cases where a single line from a web page may have been directly incorporated, which could explain subtle error propagation due to the reuse of specific code fragments.
By integrating these complementary measures across both granularities, \techname achieves a comprehensive and robust assessment of web page utilization in the generated code, ensuring that both overt and subtle cases of code reuse are effectively identified.

\noindent \textbf{Specification/Implementation Diagnosis.}
After obtaining the utilized code, \techname conducts Specification Alignment and Implementation Correctness check to diagnose the root cause.

\noindent \textit{Specification Alignment Check.}
In the specification alignment check, \techname identifies the problem addressed by each utilized code snippet and determines whether it matches the input task. This is accomplished using an LLM agent guided by an expert-designed prompt~\cite{xu2023expertprompting}, instructing the model to act as a software engineering expert specializing in problem analysis.

Guided by an expert prompt, the LLM executes a structured analytical process. First, it summarizes the requirements for the utilized web code snippet. The inputs to the LLM agent consist of the corresponding textual description and the code snippets. This summarization focuses on key aspects such as input/output formats, operational constraints (e.g., data ranges), and the primary functional goal.

Next, the LLM receives the input programming problem, including the description, a canonical correct code solution, and a set of corresponding test cases.
It compares these summarized requirements to determine if the input problem and the searched code address the same task. This comparison involves evaluating the congruence of their respective inputs, outputs, constraints, potential edge cases, and the implicit solution approach. 
Finally, the LLM determines equivalence and structures its output, which is a \textit{JSON} object adhering to a predefined schema, including \textit{WebPageProblemUseConditions} and \textit{Equivalence} keys. 
If the LLM deems the problem equivalent, it signifies a strong alignment, and vice versa. The \textit{WebPageProblemUseConditions} field provides a concise description of conditions under which the web page content can be referenced. It should be noted that when judging the requirements of web page code snippets, we do not rely on the input question requirements. Since the web page exists independently of the input question, this design method can ensure that even if the web page is used by multiple questions, the output of the large model for the web page requirements will not change.

\noindent \textit{Implementation Correctness Check.}
For web pages whose code snippets are deemed aligned with the target problem's specification, \techname proceeds to evaluate the correctness of the code snippet. 
The evaluation is conducted by leveraging the service provider's existing Correctness Evaluator. This evaluator serves as the ground truth for determining functional correctness and can take various forms depending on the provider's infrastructure—such as test suites, reference implementations, or formal specifications. 
\techname treats the evaluator as a black box and queries it for a binary correctness result. A snippet is classified as correct if it passes the evaluation, and incorrect otherwise. Any web page that fails either check is flagged as an EIP, reported to the service provider, and cached for subsequent repair.

\subsection{Repairing Phase}
Once EIPs are identified and diagnosed by the debugging phase, \techname initiates the repairing phase on the cache of these EIPs. This phase aims to rectify the diagnosed issues, transforming diagnosed EIPs into reliable content for future LLM queries. \techname employs two distinct strategies tailored to the nature of EIPs.

\noindent \textbf{Correcting flawed implementations.}
If a web page aligns with the input problem specification but contains an erroneous or buggy code snippet, \techname adopts a highly pragmatic and reliable strategy. 
It deliberately prioritizes using the evaluated, correct baseline solution as the replacement for the flawed code snippet when operating the EIP cache. This design choice is critical for our automated framework to guarantee the correctness and quality of the repair without introducing the uncertainty of generating a new solution. Furthermore, it enables a fully automated, end-to-end pipeline where SIIs can be detected and neutralized at scale, which is paramount for the LLM service providers. Considering the data labeling capabilities of large companies, this correct implementation can also be completely replaced by a manually validated answer.

\noindent \textbf{Contextualizing misaligned specifications.}
If a web page does not match the input problem specification, directly modifying its content is inappropriate, since the page may still be correct and useful in its original context. Instead, \techname injects structured metadata tags into the internally cached page representation. Derived from the debugging phase (i.e., \textit{WebPageProblemUseConditions}), these tags describe the page’s actual scope, applicable input types, algorithmic approach, and intended problem. 
This metadata helps downstream retrieval and generation components interpret the page correctly and avoid applying it to unrelated tasks. 
For LLM service providers, how to best integrate such corrected caches into their private production web systems is an important but orthogonal deployment issue that we leave to future work.

\begin{promptbox}
\footnotesize
\setlength{\tabcolsep}{2pt}
\renewcommand{\arraystretch}{0.95}
\begin{tabular}{@{}p{1.3cm}@{\hspace{0.3em}}p{\dimexpr\linewidth-2.0cm-0.3em\relax}@{}}
\textbf{Title:} & python - Efficient way to find top K products given two lists \\
\textbf{Link:} & Stack Overflow page for the queried programming problem \\
\textbf{Snippet:} & [Detected Code Snippet] \\
\textbf{Diagnosis:} & [EIP Type: Implementation Incorrectness] \\
\textbf{Time:} & [Access Time] \\
\textbf{Content:} & [Replaced Content] \\
\end{tabular}
\end{promptbox}

The repair effectiveness of \techname can be assessed by evaluating the correctness of subsequent code generations on the original problem through the ``Correctness Evaluator''. Importantly, each repair action in \techname is a one-time intervention. Once an EIP is repaired through code replacement or metadata injection, the updated version is stored internally. All future LLM queries referencing that URL are then served the rectified content, thereby neutralizing persistent threats and eliminating recurrent errors from the same web source. 
We illustrate the repaired cache corresponding to the web page depicted in Fig.\ref{PSExample}.a, which contains an incorrect implementation. In this example, \techname replaces the erroneous code snippet with the correct solution generated by the base LLM, as determined by the diagnosis phase, while leaving all other page content unchanged. Additionally, the key attributes of the EIP are preserved for internal record-keeping.
We log timestamps for each cached repair, considering that the cached repair might become outdated if the corresponding web page changes. In deployment, these can be compared against the latest web page versions to trigger re-fetching, re-validation, and, if needed, automatic cache refresh. This refresh mechanism is fully automatable. Additionally, our framework supports separate cache libraries for different task types or domains in practical deployments. This isolation ensures that repaired content tailored for one task category does not inadvertently affect unrelated tasks. 
This design enhances the long-term trustworthiness of web-augmented LLM systems.

\section{Evaluation}
\label{sec:evalution}
Our evaluation is designed to demonstrate the effectiveness of \techname. As outlined in our approach, \techname requires a Correctness Evaluator to operate. For the experiments conducted in this paper, we instantiate this evaluator using the comprehensive test suites provided by established public benchmarks. These test suites serve as a concrete and reproducible proxy for the internal quality assurance mechanisms of an LLM service provider. 
Our evaluation is structured to answer the following key research questions.

\noindent \textbf{RQ1: Debugging Efficacy.}
To what extent can \techname's debugging phase accurately detect and diagnose EIPs?
We evaluate \techname on outputs from web-augmented LLMs across five code generation/completion benchmarks, and report precision, recall, and F1-score for EIP detection and diagnosis.

\noindent \textbf{RQ2: Repairing Effectiveness.}
How effectively can \techname repair EIPs and recover erroneous LLM-generated code?
We apply \techname to SII cases where incorrect generations are caused by EIPs identified in RQ1, and measure the proportion of affected cases that are corrected after cache repair.

\noindent \textbf{RQ3: Efficiency.}
What is the computational overhead when deploying \techname in a web-augmented LLM code generation pipeline?
To answer RQ3, we analyze \techname's efficiency, which helps determine the practical viability in real-world deployment scenarios.

\noindent \textit{\textbf{Evaluated Benchmark.}}
Our evaluation includes five benchmarks covering both Python and Java.
We use \textit{HumanEval+}~\cite{evalplus}, \textit{MBPP+}~\cite{evalplus}, and \textit{HumanEval-Java}~\cite{zheng2023codegeex} as widely used code generation benchmarks; \textit{LiveCodeBench}~\cite{jain2024livecodebench} to reduce memorization concerns; and \textit{DevEval}~\cite{li2024deveval} to assess performance in more realistic repository-level development scenarios.

\noindent \textit{\textbf{Baselines.}}
To the best of our knowledge, no prior work has been specifically designed to mitigate SII through proactive cache-level repair. Therefore, there is no existing baseline that matches \techname exactly in both problem setting and intervention point. To provide a meaningful reference in RQ2, we compare \techname with the closest line of related methods, namely reactive RAG-based mitigation approaches. We choose RAG-based methods because, like \techname, they aim to reduce errors caused by imperfect external knowledge, but do so at inference time through retrieval-time or generation-time correction rather than source-level repair. Specifically, we include RA-RAG~\cite{hwang2025retrieval}, Self-RAG~\cite{asai2024self}, and RankGPT~\cite{sun2023chatgpt} as representative baselines, as they cover three common reactive mechanisms: reliability-aware retrieval, self-reflective generation, and LLM-based re-ranking. 
Since these methods intervene at a different stage of the pipeline, our comparison should be viewed not as a strict apples-to-apples competition, but as an informative comparison between proactive source-level repair and representative reactive mitigation strategies under a common evaluation protocol.

\noindent \textit{\textbf{Evaluation Metric.}}
To measure the detection and diagnosis performance of \techname, we follow previous studies~\cite{guographcodebert, wang2021codet5} and use some standard metrics in our evaluation (i.e., precision, recall, and F1-score~\cite{sparckjones1972statistical, van1979information}). 
The metrics are defined as follows: 
$
\text{Recall} = \frac{\text{TP}}{\text{TP+FN}}  ,
\text{Precision} = \frac{\text{TP}}{\text{TP+FP}},$ where TP is the number of EIPs that are classified as error-inducing; FN is the number of EIPs that are classified as correct; FP denotes the number of correct pages that are identified as error-inducing.

\noindent \textit{\textbf{Implementation.}}
Due to permission and security ethics issues, we assume that LLM service providers cannot directly modify the web content or upload incorrect code, but have the ability to cache and modify cached results. Instead, we utilize the BeautifulSoup package~\cite{beautifulsoup4} to extract the content from web pages legally. We utilize the difflib~\cite{python-difflib}, Python AST~\cite{python-ast}, and CodeBLEU~\cite{ren2020codebleu} to implement the EIP Detection. To support JSON format outputs, the specification alignment check is implemented based on LangChain~\cite{Chase_LangChain_2022} and OpenAI~\cite{openai2025python} package. We use the \texttt{GPT-4o-1120} model~\cite{gpt4o} to generate the expert prompt and act as the agent to finish judging specification misalignment. To judge the correctness of the code in the web page, we utilize the same code execution pipeline as EvalPlus and Livecodebench. Since function calling~\cite{openai-tools} is a standard strategy to supply external information in LangChain and LLM providers~\cite{o4-mini,deepseek-v3,gemini-2.5}, we follow it to provide web content to LLMs. 
The web content is prepended as optional context, and LLMs can decide whether or how to incorporate it.

\section{Results and Analysis}

\subsection{RQ1: Debugging Efficacy}
This RQ evaluates whether \techname can accurately debug EIPs in an LLM service setting.
Specifically, we assess two sequential capabilities of its debugging phase. First, we evaluate EIP detection, i.e., whether \techname can correctly identify the web pages responsible for the erroneous generations among the retrieved candidates. Second, we evaluate root-cause diagnosis, i.e., whether \techname can correctly determine whether the identified EIP causes the failure due to specification misalignment or implementation error. To answer this RQ, we construct annotated evaluation sets from the SII cases and report the performance of \techname on both detection and diagnosis using standard effectiveness metrics.

To create a dataset for evaluating \techname's debugging performance, we first collect instances where web-augmented LLMs produce incorrect code. Consistent with our preliminary study (Section~\ref{sec:study}), we select errors made by \texttt{GPT-4o} as a representative web-augmented LLM, which we further manually label to ensure that the ground truth for our evaluation is as accurate as possible.

We collect all searched web pages during \texttt{GPT-4o}'s generation process for these error cases across five benchmarks. Specifically, \texttt{GPT-4o} produces 11, 25, 10, 114, and 185 incorrect generations attributable to web content on \textit{HumanEval+}, \textit{MBPP+}, \textit{HumanEval-Java}, \textit{LiveCodeBench}, and repo-level \textit{DevEval}, respectively. For \textit{HumanEval+}, \textit{MBPP+}, and \textit{HumanEval-Java}, we analyze all web pages searched by \texttt{GPT-4o} in the context of its erroneous generations. For \textit{LiveCodeBench} and \textit{DevEval}, given the larger number of errors, we randomly sample 75 web pages that are across 25 erroneous generation instances by \texttt{GPT-4o} for each benchmark.

Two authors then manually annotate each of the collected web pages. Following the classification criteria established in Section~\ref{sec:study}, pages are labeled as either error-inducing or correct.
The inter-rater reliability for this annotation process, calculated using Cohen’s Kappa~\cite{cohen1960coefficient}, is 0.981. This high level of agreement underscores the consistency and reliability of our ground truth labels. This manually annotated dataset provides the ground truth for evaluating \techname's automated detection and diagnosis capabilities.

We deploy \techname's debugging phase to detect and diagnose each page, and then compare \techname's classifications against our manual annotations to calculate performance metrics.
The detection performance is presented in Table~\ref{tab:RQ1}. The table shows the distribution of manually labeled correct pages and error-inducing pages for each benchmark, alongside \techname's detection metrics.

Across all five benchmarks, \techname achieves strong and robust EIP detection, with F1-scores consistently above 93.00\%. Precision remains high on all benchmarks, reaching 96.43\% on \textit{HumanEval+}, 98.15\% on \textit{MBPP+}, 95.83\% on \textit{HumanEval-Java}, 96.49\% on \textit{LiveCodeBench}, and 96.97\% on \textit{DevEval}, indicating few false alarms. Recall is also consistently strong, showing that \techname identifies most true EIPs. The slightly better performance on \textit{DevEval} may be due to its repository-level setting, where richer project context makes misleading pages easier to distinguish.

\begin{table}[t!]
    \caption{Efficacy of \techname's EIP detection.}
    \label{tab:RQ1}
    \scriptsize
  
    \centering
    \scriptsize
    \begin{tabular}{l|cc|ccc}
    \toprule
    Benchmark  & Correct pages & EIPs   & F1-score & Precision & Recall \\ \midrule
    HumanEval+   & 3 & 30  &  93.10\%  &  96.43\%   &   90.00\%      \\ 
    MBPP+  & 16 & 59 &  93.81\%  & 98.15\%    &  89.83\%          \\
    HumanEval-Java & 5 & 25 & 93.88\% & 95.83\% & 92.00\% \\
    LiveCodeBench  & 15 & 60    &     94.02\%    &   96.49\%       &   91.67\%   \\    
    DevEval & 7  & 68  & 95.52\%  &  96.97\%  &  94.12\%  \\
     \bottomrule
    \end{tabular}
\end{table}

After evaluating whether \techname can detect EIPs, we next assess whether it can correctly diagnose the root cause of each identified EIP. Based on manual annotation, the 242 ground-truth EIPs consist of 205 pages with specification misalignment and 37 pages with incorrect implementations. 
\techname diagnoses these two failure types using two complementary modules: its LLM-based agent analyzes specification alignment, while its dynamic execution module checks implementation correctness. For specification diagnosis, the agent correctly identifies 188 of the 205 misaligned pages, achieving 91.71\% accuracy. For implementation diagnosis, the execution-based module correctly identifies 32 of the 37 incorrect implementations, achieving 86.49\% accuracy. Overall, \techname achieves a weighted diagnostic accuracy~\cite{bishop2006pattern} of 90.91\%.

The consistent performance across benchmarks, which vary in programming language, problem complexity, and style, suggests that \techname's detection and diagnosis mechanisms are generalizable. The ability to accurately detect and diagnose EIPs is a critical first step in mitigating its negative impact, directly enabling the subsequent repair phase and enhancing the overall reliability of web-augmented LLM systems. These results affirm the efficacy of \techname's debugging phase in addressing the challenge of SII.

\subsection{RQ2: Repairing Effectiveness}

In RQ2, we systematically evaluate the effectiveness of \techname in repairing detected EIPs. In RQ1, among the 242 manually labeled EIPs, \techname correctly detects and diagnoses 220 cases. These 220 correctly diagnosed EIPs consist of 188 cases of specification misalignment and 32 cases of incorrect code. We then assess whether the corresponding cache modifications produced by \techname correctly repair these diagnosed EIPs through detailed manual review.
A modification is considered correct if it resolves the identified issue, i.e., by accurately annotating a misaligned page with precise metadata or by replacing an incorrect code snippet with a correct solution that passes all relevant test cases. To ensure reliability, both annotators independently review every modification, resulting in a high inter-rater reliability (Cohen’s Kappa~\cite{cohen1960coefficient}) of 0.962.

Quantitatively, \techname successfully repairs 206 out of these 220 correctly diagnosed EIPs, corresponding to a repair accuracy of 93.64\%.  All incorrect repairs arise from subtle or ambiguous specification mismatches, where the boundary between relevant and misleading content is often unclear. This highlights the difficulty of precisely aligning retrieved content with user intent.

Additionally, our analysis finds that advanced LLMs are not uniformly vulnerable to misleading web content. LLMs may prevent themselves from generating errors, even in the presence of EIPs or with an incorrect modification cache. Therefore, it may be more important for LLM service providers to provide users with the correct code solutions after \techname processing. Hence, we further utilize each LLM to generate code solutions across five evaluated benchmarks under both the standard (no web search) and web-augmented settings. To rigorously control for model stochasticity, we follow the correctness decision criteria described in Section~\ref{sec:study} to identify SII cases. For each identified SII case, we collect the corresponding searched web pages and apply our \techname framework to debug and repair the problematic content. Finally, we use the benchmark's test suite (our instantiated Correctness Evaluator) to check the correctness of code generation after \techname's processing.

\begin{table}[t]
\centering
\caption{\techname's repair effectiveness across benchmarks.}
\label{tab:RQ2}

\scriptsize
\setlength{\tabcolsep}{0.5pt}
\begin{tabular}{lcccccccccc}
\toprule
\textbf{Model Name} & 
\multicolumn{2}{c}{\textit{HumanEval+}} & 
\multicolumn{2}{c}{\textit{MBPP+}} &
\multicolumn{2}{c}{\textit{HumanEval-Java}} & 
\multicolumn{2}{c}{\textit{LiveCodeBench}} &
\multicolumn{2}{c}{\textit{DevEval}}\\
\cmidrule(lr){2-3} \cmidrule(lr){4-5} \cmidrule(lr){6-7} \cmidrule(lr){8-9} \cmidrule(lr){10-11}
& \textbf{Nums} & \textbf{Repair} & \textbf{Nums} & \textbf{Repair} & \textbf{Nums} & \textbf{Repair} & \textbf{Nums} & \textbf{Repair} & \textbf{Nums} & \textbf{Repair}\\
\midrule
\texttt{GPT-4o} & 11 & 10 & 25 & 24 & 10 & 10 & 114 & 99 & 185 & 171\\
\texttt{DeepSeek-V3} & 15 & 13 & 16 & 14 & 12 & 11 & 72 & 53 & 156 & 133\\
\texttt{Claude 3.7 Sonnet} & 7 & 7 & 11 & 10 & 4 & 3 & 33 & 24 & 94 & 78\\
\texttt{Gemini 2.5 Flash} & 14 & 13 & 25 & 22 & 6 & 5 & 79 & 58 & 159 & 142\\
\texttt{o4-mini} & 7 & 5 & 44 & 41 & 3 & 3 & 59 & 51 & 124 & 116\\
\texttt{DeepSeek-R1} & 8 & 7 & 15 & 14 & 7 & 6 & 67 & 52 & 147 & 121\\
\bottomrule
\end{tabular}
\end{table}

\begin{table*}[t!]
\centering
\caption{Comparison of repair effectiveness among \techname and baselines under Top-3, Top-5, and Top-10 retrieval settings.}
\label{tab:rq2_rag_compare_new}
\setlength{\tabcolsep}{1.5pt}
\renewcommand{\arraystretch}{1.05}
\footnotesize  
\begin{tabular}{llccccccccccccccc}
\toprule
\multirow{2}{*}{\textbf{Model}} & \multirow{2}{*}{\textbf{Method}} 
& \multicolumn{3}{c}{\textit{HumanEval+}} 
& \multicolumn{3}{c}{\textit{MBPP+}}
& \multicolumn{3}{c}{\textit{HumanEval-Java}} 
& \multicolumn{3}{c}{\textit{LiveCodeBench}}
& \multicolumn{3}{c}{\textit{DevEval}} \\
\cmidrule(lr){3-5} \cmidrule(lr){6-8} \cmidrule(lr){9-11} \cmidrule(lr){12-14} \cmidrule(lr){15-17}
& 
& \textbf{Top-3} & \textbf{Top-5} & \textbf{Top-10}
& \textbf{Top-3} & \textbf{Top-5} & \textbf{Top-10}
& \textbf{Top-3} & \textbf{Top-5} & \textbf{Top-10}
& \textbf{Top-3} & \textbf{Top-5} & \textbf{Top-10}
& \textbf{Top-3} & \textbf{Top-5} & \textbf{Top-10} \\
\midrule

\multirow{4}{*}{\texttt{GPT-4o}}
& \techname   & 90.91\% & 92.31\% & 76.47\% & 96.00\% & 92.59\% & 82.35\% & 100.00\% & 91.67\% & 78.57\% & 86.84\% & 85.25\% & 81.06\% & 92.43\% & 92.19\% & 89.16\% \\
& RA-RAG      & 45.45\% & 46.15\% & 47.06\% & 52.00\% & 51.85\% & 47.06\% & 40.00\% & 41.67\% & 42.86\% & 45.61\% & 46.72\% & 44.70\% & 47.03\% & 46.35\% & 45.32\% \\
& Self-RAG    & 27.27\% & 23.08\% & 23.53\% & 28.00\% & 29.63\% & 26.47\% & 30.00\% & 25.00\% & 28.57\% & 24.56\% & 24.59\% & 23.48\% & 24.86\% & 25.00\% & 24.14\% \\
& RankGPT     & 9.09\% & 7.69\% & 5.88\% & 12.00\% & 11.11\% & 11.76\% & 10.00\% & 8.33\% & 7.14\% & 4.39\% & 4.92\% & 6.82\% & 8.65\% & 8.85\% & 8.37\% \\
\midrule

\multirow{4}{*}{\texttt{DeepSeek-V3}}
& \techname   & 86.67\% & 88.24\% & 84.21\% & 87.50\% & 87.50\% & 83.33\% & 91.67\% & 91.67\% & 85.71\% & 73.61\% & 72.73\% & 72.29\% & 85.26\% & 85.28\% & 80.79\% \\
& RA-RAG      & 40.00\% & 47.06\% & 42.11\% & 43.75\% & 43.75\% & 44.44\% & 41.67\% & 50.00\% & 42.86\% & 47.22\% & 46.75\% & 48.19\% & 45.51\% & 44.79\% & 43.50\% \\
& Self-RAG    & 26.67\% & 23.53\% & 26.32\% & 25.00\% & 18.75\% & 16.67\% & 33.33\% & 33.33\% & 28.57\% & 22.22\% & 22.08\% & 24.10\% & 24.36\% & 25.15\% & 24.29\% \\
& RankGPT     & 6.67\% & 11.76\% & 10.53\% & 6.25\% & 6.25\% & 5.56\% & 8.33\% & 8.33\% & 7.14\% & 6.94\% & 6.49\% & 7.23\% & 5.77\% & 5.52\% & 6.21\% \\
\midrule

\multirow{4}{*}{\texttt{Claude 3.7}}
& \techname   & 100.00\% & 100.00\% & 88.89\% & 90.91\% & 91.67\% & 86.67\% & 75.00\% & 80.00\% & 75.00\% & 72.73\% & 74.29\% & 71.05\% & 82.98\% & 83.51\% & 80.77\% \\
& RA-RAG      & 57.14\% & 57.14\% & 44.44\% & 54.55\% & 50.00\% & 46.67\% & 50.00\% & 40.00\% & 37.50\% & 45.45\% & 45.71\% & 44.74\% & 47.87\% & 46.39\% & 46.15\% \\
& Self-RAG    & 28.57\% & 42.86\% & 33.33\% & 36.36\% & 33.33\% & 33.33\% & 25.00\% & 20.00\% & 25.00\% & 27.27\% & 28.57\% & 26.32\% & 32.98\% & 34.02\% & 33.65\% \\
& RankGPT     & 14.29\% & 14.29\% & 11.11\% & 9.09\% & 8.33\% & 6.67\% & 0.00\% & 0.00\% & 12.50\% & 6.06\% & 8.57\% & 7.89\% & 6.38\% & 7.22\% & 7.69\% \\
\midrule

\multirow{4}{*}{\texttt{Gemini 2.5}}
& \techname   & 92.86\% & 93.33\% & 84.21\% & 88.00\% & 86.21\% & 81.82\% & 83.33\% & 85.71\% & 83.33\% & 73.42\% & 72.41\% & 74.04\% & 89.31\% & 89.09\% & 84.41\% \\
& RA-RAG      & 42.86\% & 40.00\% & 47.37\% & 44.00\% & 37.93\% & 39.39\% & 50.00\% & 42.86\% & 41.67\% & 44.30\% & 44.83\% & 42.31\% & 45.91\% & 47.27\% & 45.70\% \\
& Self-RAG    & 35.71\% & 33.33\% & 31.58\% & 24.00\% & 20.69\% & 21.21\% & 16.67\% & 28.57\% & 25.00\% & 26.58\% & 27.59\% & 27.88\% & 29.56\% & 29.70\% & 28.49\% \\
& RankGPT     & 7.14\% & 6.67\% & 5.26\% & 8.00\% & 6.90\% & 6.06\% & 0.00\% & 14.29\% & 8.33\% & 5.06\% & 5.75\% & 6.73\% & 6.92\% & 7.27\% & 6.45\% \\
\midrule

\multirow{4}{*}{\texttt{o4-mini}}
& \techname   & 71.43\% & 75.00\% & 70.00\% & 93.18\% & 92.16\% & 87.30\% & 100.00\% & 100.00\% & 83.33\% & 86.44\% & 86.36\% & 85.00\% & 93.55\% & 92.42\% & 90.60\% \\
& RA-RAG      & 42.86\% & 50.00\% & 40.00\% & 43.18\% & 47.06\% & 44.44\% & 33.33\% & 33.33\% & 16.67\% & 40.68\% & 43.94\% & 46.25\% & 45.97\% & 46.97\% & 46.31\% \\
& Self-RAG    & 14.29\% & 25.00\% & 20.00\% & 22.73\% & 21.57\% & 20.63\% & 0.00\% & 0.00\% & 16.67\% & 28.81\% & 27.27\% & 32.50\% & 25.81\% & 28.79\% & 30.20\% \\
& RankGPT     & 0.00\% & 12.50\% & 10.00\% & 6.82\% & 7.84\% & 7.94\% & 0.00\% & 0.00\% & 16.67\% & 6.78\% & 6.06\% & 6.25\% & 6.45\% & 6.06\% & 6.71\% \\
\midrule

\multirow{4}{*}{\texttt{DeepSeek-R1}}
& \techname   & 87.50\% & 88.89\% & 83.33\% & 93.33\% & 94.12\% & 85.71\% & 85.71\% & 85.71\% & 80.00\% & 77.61\% & 77.78\% & 77.65\% & 82.31\% & 82.39\% & 81.98\% \\
& RA-RAG      & 62.50\% & 55.56\% & 41.67\% & 46.67\% & 41.18\% & 38.10\% & 57.14\% & 57.14\% & 50.00\% & 50.75\% & 48.61\% & 49.41\% & 46.94\% & 46.54\% & 51.74\% \\
& Self-RAG    & 37.50\% & 33.33\% & 33.33\% & 33.33\% & 35.29\% & 28.57\% & 42.86\% & 42.86\% & 40.00\% & 37.31\% & 37.50\% & 32.94\% & 31.97\% & 30.82\% & 30.81\% \\
& RankGPT     & 12.50\% & 11.11\% & 8.33\% & 6.67\% & 5.88\% & 9.52\% & 14.29\% & 14.29\% & 10.00\% & 5.97\% & 5.56\% & 7.06\% & 7.48\% & 6.92\% & 6.98\% \\

\bottomrule
\end{tabular}
\end{table*}

Table~\ref{tab:RQ2} presents \techname’s repair performance. For each evaluated code generation model and benchmark, the Nums column records the number of tasks affected by SII, while the Repair column reports how many of these tasks produce correct solutions after \techname debugs and repairs the implicated web pages. Across all evaluated models and datasets, \techname rectifies the majority of SII cases. For example, for \texttt{GPT-4o}, it repairs 10/11 cases on \textit{HumanEval+}, 24/25 on \textit{MBPP+}, 10/10 on \textit{HumanEval-Java}, 99/114 on \textit{LiveCodeBench}, and 171/185 on \textit{DevEval}. Similar trends hold for other models.  
\techname remains highly effective even on unleaked \textit{LiveCodeBench} and repo-level \textit{DevEval}, suggesting that its benefits are not limited to simpler benchmark settings but also generalize to more realistic and structurally complex programming tasks.

\subsubsection{Safety and potential negative side effects.}
Since \techname applies a one-time correction to each implicated web page and stores the modified version in a cache, all subsequent tasks referencing that page will utilize the updated content. 
A critical consideration for a proactive framework like \techname is ensuring that repairs intended to fix one problem do not inadvertently introduce regressions for others. An incorrect EIP classification or a repair could poison the cache for other tasks. 

To evaluate this risk, we design an experiment to measure the potential negative side effects of \techname's repairs. For each web page cache repaired by \techname, we identify all programming problems for which the original page yields correct solutions. After repair, we regenerate solutions for these programming problems using the modified cache. To balance comprehensiveness and computational efficiency, this analysis is conducted on four representative LLMs: \texttt{GPT-4o}, \texttt{DeepSeek-V3}, \texttt{o4-mini}, and \texttt{DeepSeek-R1}.

Across the five benchmarks, we identify 107 pages that are referenced by multiple programming problems.
Of these, \techname repairs 83 pages, which collectively appear in 96 other tasks where the original code generation is correct. For each of these 96 cases, we generate new solutions using the cache. Our results show that repairs introduced by \techname do not degrade the correctness of any previously unaffected solutions in evaluated benchmarks. 
In other words, fixes applied for one problem generalize safely to all other programming problems observed in our experiments. This indicates that \techname can preserve the reliability of web-augmented LLM applications in realistic deployment scenarios.

\subsubsection{Comparison of RAG baselines}
We further compare \techname with three representative reactive baselines, \textit{RA-RAG}, \textit{Self-RAG}, and \textit{RankGPT}. Table~\ref{tab:rq2_rag_compare_new} shows that \techname consistently achieves the best repair performance in all 90 evaluated model--benchmark--Top-$k$ settings. Here, the repair rate for each setting is defined as the proportion of SII-affected tasks that are corrected after applying the mitigation method. Macro-averaged over the 30 model--benchmark pairs under each retrieval depth, \techname achieves repair rates of 87.02\%, 87.02\%, and 81.63\% for Top-3, Top-5, and Top-10, respectively. In contrast, the strongest baseline, RA-RAG, reaches only 46.68\%, 46.25\%, and 43.62\%,
while Self-RAG and RankGPT obtain much lower repair rates.
This advantage also persists on more challenging benchmarks. On repository-level \textit{DevEval}, \techname achieves 87.64\%, 87.48\%, and 84.62\% under Top-3, Top-5, and Top-10, respectively, compared with 46.54\%, 46.38\%, and 46.45\% for RA-RAG. Similar trends are observed on \textit{LiveCodeBench}, suggesting that \techname remains effective in more realistic settings with longer contexts and noisier retrieval.

We also observe a clear trade-off as retrieval depth increases. Expanding retrieval from Top-3 to Top-5 does not consistently improve repair performance. For \techname, the macro-average remains 87.02\%, with gains and drops largely offsetting each other. In contrast, increasing the depth to Top-10 leads to a clearer decline. \techname drops to 81.63\%, with 26 of 30 model--benchmark pairs performing worse than under Top-3, and RA-RAG also decreases from 46.68\% to 43.62\%.
This trend suggests that deeper retrieval in open-web settings exposes models to more low-quality, outdated, or context-misaligned pages. Although retrieving more pages may occasionally introduce useful evidence, the marginal benefit quickly diminishes while the risk of additional EIPs accumulates. 

Overall, these results suggest that proactive source repair is more effective than representative reactive RAG-style mitigation for addressing SIIs, and remains robust as retrieval depth increases. This also motivates a natural next question: whether proactive repair can be further combined with reactive mitigation.

\noindent \textbf{Combining \techname with RA-RAG.}
To further examine whether proactive source repair is complementary to reactive RAG mitigation, we conduct an additional experiment on the two most challenging benchmarks, \textit{LiveCodeBench} and \textit{DevEval}, by combining \techname with the strongest reactive baseline, \textit{RA-RAG}. Specifically, we first apply \techname to diagnose and repair error-inducing pages in the cache, and then perform generation with RA-RAG over the repaired retrieval pool. Using \texttt{GPT-4o}, this combined strategy further improves repair effectiveness over either method alone. On \textit{LiveCodeBench}, the repair rate reaches 94.74\%, 94.26\%, and 93.94\% under Top-3, Top-5, and Top-10, respectively. On \textit{DevEval}, the corresponding repair rates are 95.68\%, 94.79\%, and 93.10\%. These results suggest that proactive source-level repair and reactive reliability-aware retrieval are complementary: once misleading pages are repaired at the source, downstream RAG mechanisms can make more effective use of the remaining evidence, yielding even stronger robustness on complex and noisy retrieval settings.

\subsection{RQ3: Efficiency Evaluation}

To assess \techname's practical feasibility for LLM service providers, we evaluate its time overhead for each phase of operation.
We calculate the average per-case execution time for the main components.

In the debugging phase, \techname requires an average of 2.94s per case. This consists of 0.41s for EIP detection and 2.53s for specification/implementation diagnosis. Notably, the majority of the diagnosis time (2.49s) is attributed to specification alignment checks, primarily due to the latency of LLM API calls. The implementation correctness check (test execution) incurs modest overhead, averaging only 0.04s per case. For the repairing phase, \techname demonstrates high efficiency, completing repairs in under 0.01s on average. We believe that the time cost can be further reduced by adopting parallel techniques and the deployment of LLMs.

These results indicate that \techname can be seamlessly integrated into web-augmented LLM pipelines without incurring high computational cost. 
It ensures both timely detection and correction of EIPs in practical deployment scenarios for LLM service providers.

\section{Discussion}

\noindent \textbf{Comparison with Existing Code Repair.}
Existing code repair studies~\cite{jimenezswe, just2014defects4j, xia2024automated, yin2024thinkrepair, zhu2021syntax} focus on fixing erroneous code from the developer perspective. In contrast, we address SII in web-augmented LLMs from the service-provider perspective by debugging and repairing the underlying EIPs. Unlike post-generation code repair, \techname performs proactive source-level repair, providing longer-term benefits. It may also help reduce potential pollution from erroneous web data in future LLM training.

\noindent \textbf{Potential Application.}
\techname is designed for LLM service providers to improve the reliability of web-augmented code generation. Its effectiveness depends on the quality of the ``Correctness Evaluator''; although we use benchmark test suites~\cite{zheng2023codegeex,evalplus}, real-world evaluators may be imperfect. Future work includes improving evaluator reliability through redundant validation, scaling benchmark construction to cover broader programming domains, extending \techname to multimodal web content, and integrating it with vulnerability detection and repair techniques~\cite{zhou2024large, zhang2024prompt, luo2024scvhunter} to further enhance robustness.

\begin{figure}[]
\centering
\includegraphics[width=0.98\linewidth]{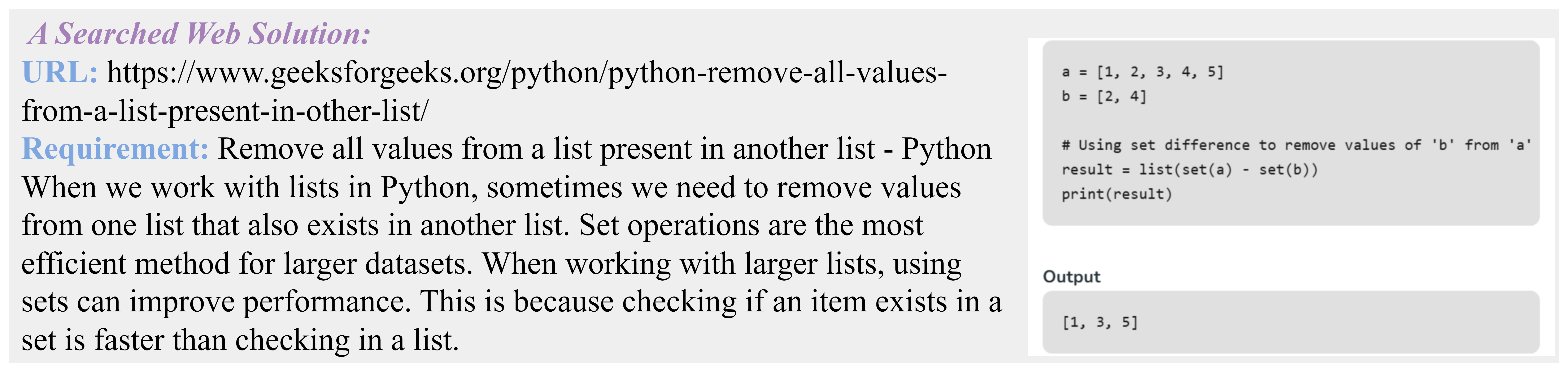}
\caption{A representative failure case of \techname.}
\label{repair_failure}
\end{figure}

\noindent \textbf{Case Study.}
Although \techname repairs most EIPs, it still fails in a few cases. Fig.~\ref{repair_failure} shows one example from \textit{MBPP+ Task 161}, which requires removing from one list all elements appearing in another, including potentially unhashable elements. A retrieved web page instead reformulates the task as removing all values and provides a set-based subtraction solution, implicitly assuming hashable elements. Because \techname mainly relies on code-level similarity, it misclassifies this page as correct and misses the underlying specification mismatch.
This case exemplifies \techname’s core limitation: it struggles with subtle or abstract discrepancies in problem intent, particularly those tied to type semantics or usage constraints not explicit in the surface code. 
Such failures highlight the need for integrating more robust semantic reasoning or formal specification checks into \techname. For example, type-aware static analysis or constraint-based reasoning could help flag violations that are not directly observable in code structure alone. Addressing these limitations represents an important direction for future work.

\noindent \textbf{Threats to Validity.}
We acknowledge the following threats.

\textit{Threats to internal validity.}
Although we include six advanced mainstream LLMs, the evaluated model set is still not exhaustive. Prompt selection may also affect the results; to reduce this threat, we directly use the released benchmark prompts. 
Manual annotation is another potential threat. To minimize it, two authors independently annotate each page and resolve disagreements through discussion, achieving high agreement ($> 0.9$)~\cite{cohen1960coefficient}.
We also acknowledge that our emulated search protocol may not fully match the dynamic retrieval and search-triggering strategies used in commercial systems. In particular, we assume that search is triggered for every code generation task, which may overestimate the real-world prevalence of SII. Therefore, the reported NIR should be interpreted as evidence of the existence and characteristics of SII under search-activated settings, rather than as an exact estimate of its prevalence in deployed systems. To partially mitigate this threat, we include three real web-augmented LLM systems in the preliminary study, confirming that SII is a practical phenomenon. Moreover, since \techname repairs EIPs at the cache level, its applicability is less dependent on the specific search pipeline or trigger logic.

\textit{Threats to external validity.} The utilized dataset may influence the generalization of our findings. To mitigate this risk, we employ EvalPlus~\cite{evalplus} (including HumanEval+ and MBPP+), HumanEval-Java~\cite{zheng2023codegeex}, LiveCodeBench~\cite{jain2024livecodebench}, and DevEval~\cite{li2024deveval}, which are all widely used datasets. However, due to the high cost of utilizing LLMs, our evaluation only involves code generation and completion. The generalization of our findings to other tasks remains uncertain. Future work will involve more extensive evaluations across various tasks to thoroughly assess SII and EIP's generalizability.

\section{Conclusion}
In this work, we systematically investigate Search-Induced Issues (SII), a critical vulnerability for web-augmented LLMs. Our empirical study across multiple commercial and open-source models demonstrates that SII is a prevalent risk, with an average of 73\% new errors introduced for every problem fixed via web search. To address this, we introduce \techname, an automated defense framework for LLM service providers. Our experiments show that \techname can accurately detect the root cause of these issues (EIPs) with an F1-score of up to 95\% and subsequently correct 71\% to 100\% of the resulting generation failures, all while maintaining system safety by avoiding negative side effects. It significantly improves the reliability of web-augmented LLM in code generation.

\bibliographystyle{ACM-Reference-Format}
\bibliography{ref}

\end{document}